\documentclass{revtex4-1}

\usepackage{graphicx}

\begin{document}
 \title{On the maximum sufficient range of interstellar vessels}

\author{Daniel Cartin}
\email{cartin@naps.edu}\affiliation{Naval Academy Preparatory School, 197 Elliot Street, Newport, Rhode Island 02841}

\date{\today}
\begin{abstract}

This paper considers the likely maximum range of space vessels providing the basis of a mature interstellar transportation network. Using the principle of sufficiency, it is argued that this range will be less than three parsecs for the average interstellar vessel. This maximum range provides access from the Solar System to a large majority of nearby stellar systems, with total travel distances within the network not excessively greater than actual physical distance.

\end{abstract}

\maketitle

The last few decades have seen design studies -- such as Project Daedalus~\cite{ProDae}, its successor, Icarus~\cite{ProIca}, and the Longshot effort~\cite{ProLon} -- aimed to create realistic plans for future vehicles able to travel to other solar systems. These designs are meant to be the first step in developing engineering methods for the transportation needs of an interstellar civilization. Because of the great gulf between our current knowledge and the daunting nature of this task, much of this effort has focused on developing the appropriate technology for travel between the stars within a reasonable amount of time with a large probability of success. However, it is interesting to consider further into the future, when engineering expertise is developed to make interstellar travel relatively commonplace. Specifically, the question is what are the likely characteristics of the products of a mature interstellar travel technology. In this work, it is argued that the maximum travel distance capability (before resupply at a port of call) of a typical interstellar vehicle will be on the order of three parsecs (pc). Although technological development may continue to progress, increasing the maximum possible travel distance, the factors of cost and sufficiency will conspire to limit the need for craft with longer ranges.

Many assume that as interstellar travel becomes more frequent, the range of such spacecraft will increase hand-in-hand with improving technologies and designs. An example of this is the idea of the ``incessant obsolescence postulate"~\cite{Gli04} or the "incentive trap"~\cite{Ken06}, a suggestion that the earliest starships launched towards another star will actually arrive later than subsequent missions due to improvements in speed and other characteristics. To some extent this is in line with the development of transportation technologies on the Earth's surface -- for example, the increasing range of aircraft over the last century. However, this avoids the question of sufficiency, as opposed to ability. To continue the example of aircraft, it is certainly in the realm of possibility to design airplanes with global ranges, but this is not done for several reasons. First, these craft would have a vastly increased resource cost over current designs. In addition, it has been deemed better to operate aircraft with shorter ranges, to match a transportation system that caters to passengers with a wide range of destinations. This type of design can minimize the various costs in the complete transportation system (see, e.g.~\cite{TayWec}).

Specifically for interstellar spacecraft, there are other considerations to factor into the effective safe travel distance of a vessel. One is the need for self-repairing systems; since these craft will be isolated for years, if not decades, any damage must be repaired within the vessel, rather than relying on outside help. It is likely that there is a roughly fixed chance of equipment failure happening for a given amount of time. Thus, extending the travel range of a vehicle without an accompanying change in cruise speed results in the increasing probability of either a single critical fault or multiple such incidents resulting in collapse of vehicle effectiveness and mission failure. Another issue is the amount of shielding required to protect sensitive passengers and control equipment -- as the cruise velocity of a vessel increases (and thus the range), so does the amount of radiation and collision protection required in place. Technological advances may result in active, rather than passive, means to shield the craft (e.g. magnetic fields vice large amounts of "dumb" material in front of vital areas), but this returns again to the worry about critical failures of active shielding mechanisms. Finally, there is the simple matter of resource and effort costs. Although longer range vessels may be possible, if there is insufficient need for them or ship cargos are directed to multiple destinations, the same resources can be devoted to constructing a greater number of shorter range craft to facilitate the transportation network.

Thus, in this work we consider how the properties of an interstellar transportation network vary with the maximum physical distance $D_{max}$ of its links between stellar systems. First, the total number of star systems reachable from the Solar System changes is examined, setting the lower limit for $D_{max}$. After this is done, a comparison is made between the travel distance within the transportation network -- where the maximum distance between systems is fixed -- compared to the actual physical distance between the departure and destination systems, for trips starting at the Solar System. In particular, how the ratio of these two distances, geodesic distance inside the transportation network to physical distance, changes as the spacecraft range $D_{max}$ is increased. The requirement that the average value of this ratio is not ``too large" will set the largest sufficient value for $D_{max}$. Note the emphasis on maximum range, rather than on an upper time in travel time. Presumably such times will decrease as technology advances, but the range of travel distance is predicated solely on the already existing physical distances between nearby stars.

Before presenting the results, the method of deciding on the maximum travel range is briefly discussed. This paper depends crucially on the principle of sufficiency for the transportation network -- namely, what is the largest travel range needed to connect most of the star systems in the Solar neighborhood, without requiring vessels to execute journeys of distances much greater than the actual physical distance. A more rigorous analysis -- using linear optimization, for example, to minimize the total resource cost of the transportation network -- would require an engineering model to describe the additional resources needed to improve spacecraft range capabilities by a specific amount. If this model is acceptable, an optimization procedure could design the most efficient network, given transportation supply and demand at each stellar system. This would include the possibility of several interstellar vehicle designs, suited to particular duties with the network. However, the issue is that these supply and demand factors depend crucially on $D_{max}$. Suppose transportation needs within the travel network are based roughly on random walks within the network. As a result, equilibrium supply and demand at a particular star system would depend on the number of connections this system makes with its neighbors~\cite{Lov96}. However, this degree increases with the travel range $D_{max}$. Even if there was a reasonable engineering model of cost versus capability, finding the optimal $D_{max}$ would be a non-linear optimization problem. Thus, we consider only an averaged range capability over all interstellar vessels that is sufficient to connect most star systems in the Solar neighborhood as a first approximation of this more optimal transportation network.

Moving on to the simpler situation described here, the term ``star system" as used here includes all large-mass objects -- not just stars (in the sense of masses shining due to nuclear fusion), but also white and brown dwarfs. The supposition here is that a civilization capable of building reliable interstellar craft will pass through an interplanetary phase first, due to the enormous energy requirements for interstellar (compared to interplanetary) colonization, developing the capability to colonize a variety of environments, such as asteroids, comets, and other non-Earth-like locations.~\cite{Dra85}. Thus, there would be no a priori bias towards certain types of stars, with spectral types close to the Sun's; instead such a civilization would be capable of developing resources and living structures across a wide range of habitats. Singular white and brown dwarfs may be important solely due to their role in the interstellar transportation network, serving as repair or replenishment stations for vehicles en route to other destinations. In addition, some stars with sufficient distance between them are counted as separate stellar systems, e.g. Proxima Centauri is treated as distinct from the $\alpha$ Centauri A/B system. Finally, for the purposes of this paper, only stellar systems within 15 pc of the Sun are considered, giving a total sample size of 689 systems, including the Solar System; the overall picture does not change appreciably for differing samples.

\begin{table}
\begin{tabular}{|c|c|c|c|}
\hline
Maximum travel distance (pc) & Largest component size	& Solar component size	& Mean component size	\\
\hline
\hline
2.0	& 75		& 5		& 2.7671		\\
\hline
2.2	& 153	& 5		& 4.0529		\\
\hline
2.4	& 399	& 399	& 5.9913		\\
\hline
2.6	& 476	& 476	& 8.6125		\\
\hline
2.8	& 562	& 562	& 13.25		\\
\hline
\end{tabular}
\caption{\label{comp-table}Variation in the sizes of the connected components in the interstellar transport network as a function of the maximum travel distance. The total sample size is $N=689$, consisting of all stellar systems within 15 pc of the Sun (including our own Solar System). Note that over 80\% of all such systems are reachable from the Solar System when $D_{max} = 2.8$ pc.}
\end{table}

The first question is what is the minimum travel distance $D_{max}$ needed for an interstellar vehicle. Obviously, $D_{max} > 1.3$ pc in order to reach the closest star, Proxima Centauri. However, unless $D_{max} > 2.2$ pc, then only a limited number of stellar systems are reachable -- Proxima Centauri as well as the two stars of the $\alpha$ Centauri system, Barnard's Star, and Ross 154. Thus, the minimum sufficient $D_{max}$ is about 2.3 pc, which allows for a connected transportation network of five systems. This is seen in Table \ref{comp-table}, listing the size of all stellar systems connected to the Sun (``Solar component") by vessels with a given $D_{max}$, compared to the largest transportation network possible for all the stars within 15 pc of the Sun. The interesting fact, however, is that this $D_{max}$ of 2.3 pc is rather large for the local stellar neighborhood. For example, with interstellar vehicles with a range $D_{max} = 2.0$ pc, there is already a group of 75 star systems that are reachable from one another, increasing to 153 for $D_{max} = 2.2$ pc. Thus, a hypothetical interstellar civilization arising in one of those other stellar systems would have less need for longer range spacecraft, which may serve as a possible reason the Solar System has apparently not been visited by such vessels.

\begin{table}
\begin{tabular}{|c|c|c|}
\hline
Maximum travel 	& Mean total		& Maximum total	\\
distance (pc) 		& distance ratio		& distance ratio		\\
\hline
\hline
2.3	& 2.8850	& 8.9655	\\
\hline
2.6	& 1.6649	& 3.4845	\\
\hline
2.9	& 1.3050	& 2.2548	\\
\hline
\end{tabular}
\caption{\label{dist-ratio}Characteristics of the ratio between the total stellar network geodesic distance and the physical distance, as a function of the maximum travel distance. All paths considered start at the Solar System; these ratios are computed for the $N = 257$ stellar systems reachable from the Solar System when $D_{max} = 2.3$ pc.}
\end{table}

\begin{figure}
	\includegraphics[width=0.75\textwidth]{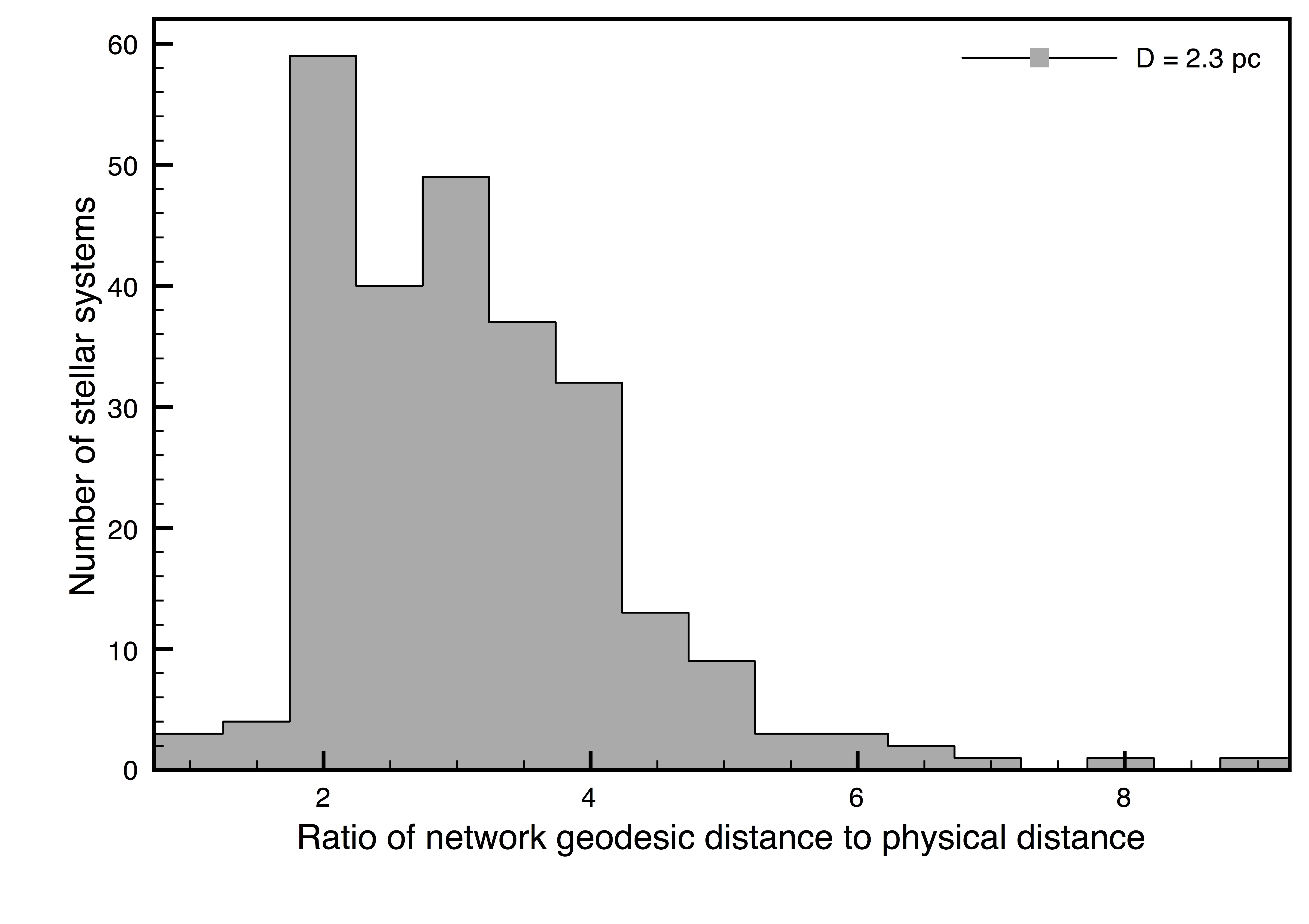}
	\caption{\label{ratios-graph-01}Histogram of the ratio of travel distance to physical distance from the Solar System for an interstellar vessel with a maximum travel range of 2.3 pc ($N = 257$ systems). Note that going from the Solar System to almost all other stellar systems requires twice the travel distance that a direct path would take.}
\end{figure}

Now that this lower limit is established, the largest sufficient range necessary is considered next. One issue is that too low of a vehicle range means that some stars are reached from the solar system by an excessively long path, through multiple other star systems. This means the ratio of the travel distance required by an interstellar vessel with a maximum range limit to the actual physical distance between the Sun and another star system may be inordinately large (although what value of the ratio is ``too large" depends on the use of such a vessel). To quantify this  aspect of the vehicle maximum range, Table \ref{dist-ratio} shows the mean and maximum distance ratios as a function of the maximum travel range. Thus, at $D_{max} = 2.3$ pc -- the minimum range to allow contact with a large number of other stellar systems -- the distance traveled by such an interstellar vessel to another star would, on average, be almost three times the straight-line distance between the Sun and that star. The variation in this travel ratio is shown in Figure \ref{ratios-graph-01}. Even worst is the maximum ratio of 8.97 for the relatively nearby star of Wolf 359; if a starship only had a maximum range of 2.3 pc, such a craft would be required to travel through multiple star systems to reach its destination.

\begin{figure}
	\includegraphics[width=0.75\textwidth]{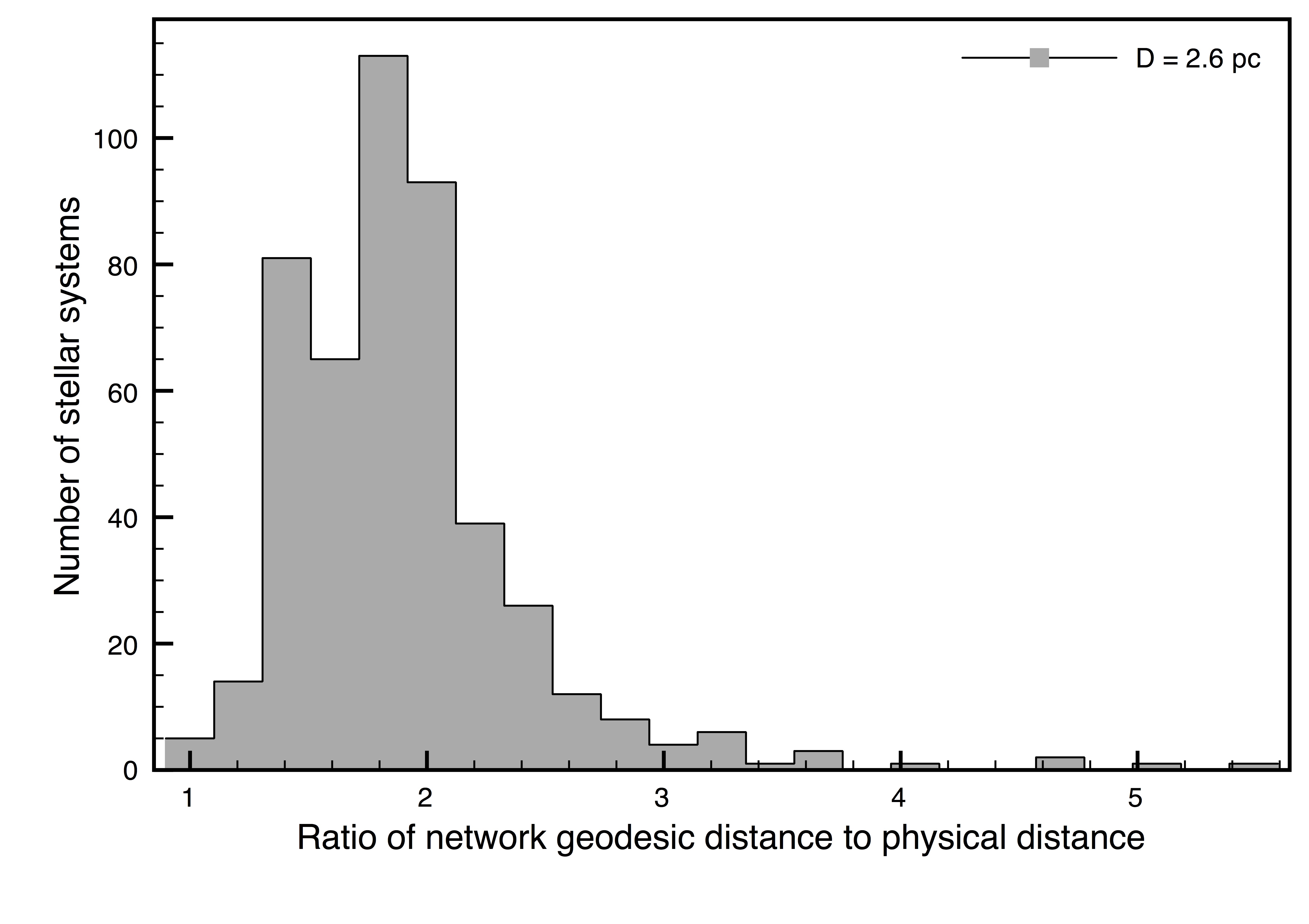}
	\includegraphics[width=0.75\textwidth]{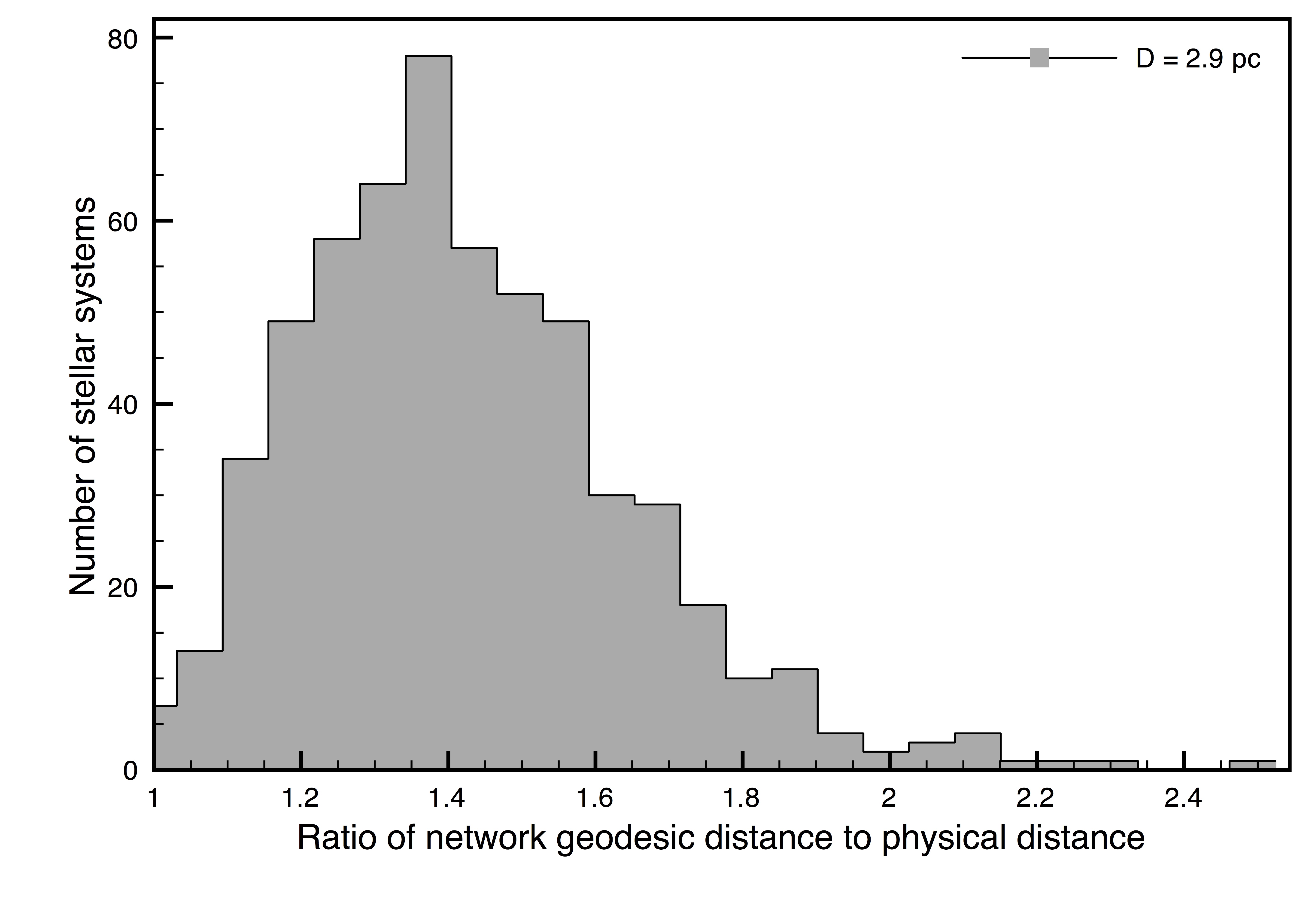}
	\caption{\label{ratios-graph-02}Histograms of the ratio of travel distance to physical distance from the Solar System for an interstellar vessel with maximum travel ranges $D_{max}$ of 2.6 pc and 2.9 pc. Unlike Table \ref{dist-ratio}, these graphs count all stellar systems reachable from the Solar System at a given $D_{max}$. When the maximum range is 2.6 pc, $N= 475$ star systems, and the average distance ratio is 1.80; when $D_{max} = 2.9$ pc, $N=576$ stellar systems and the average distance ratio is 1.39.}
\end{figure}

Expanding the maximum range $D_{max}$ of spacecraft reduces this ratio of travel to physical distance. If we compare the distance ratio for the same $N = 257$ star systems reachable when $D_{max} = 2.3$ pc, we find the average ratio decreases to 1.66 and 1.31 when $D_{max}$ is 2.6 pc and 2.9 pc, respectively, as seen in Table \ref{dist-ratio}. Therefore, by the time the maximum vessel range is 2.9 pc, the travel distance to the average stellar system from the Solar System within the transportation network is on average 31\% greater than a straight line journey. In addition, this route typically includes one or more ports of call along the way, allowing for both resupply and repair of the vessel, and facilitating connections with other vessels traveling to multiple destinations. Even when all star systems accessible with $D_{max} = 2.6$ or 2.9 pc are included, as seen in Figure \ref{ratios-graph-02}, the maximum distance ratio decreases substantially compared to the minimum travel range of 2.3 pc.

In this paper, it is argued that a mature technology will produce spacecraft with a maximum travel range of about three parsecs, based on the principle of sufficiency. Using the air transportation network on Earth as a analogy, it is likely that the corresponding interstellar version would balance minimization of the distance traveled on average by passengers or cargo with the resource cost necessary to do this. Although specialized applications -- such as courier services or military vessels -- may require greater distance ranges, the fact that a majority of star systems in the Solar neighborhood are reachable without excessive increased travel distance within the transportation network keeps such travel efficient in terms of capability versus resource usage. Designing longer range craft would be feasible, but in general would not be a good balance of cost versus need.

\end{document}